\documentclass{article}

\newcommand{\dfrac}{\displaystyle \frac }
\newcommand{\tfrac}{\textstyle \frac }

\newcommand{\prg}[1]{\mathbf {#1}}
\newcommand{\nkg}[1]{\mbox {\boldmath $#1$}}
\newcommand{\slin}[1]{\mbox {\fontsize{8pt}{1mm}\selectfont $ #1 $}}

\def\pd#1#2{\dfrac{\partial#1}{\partial#2}}

\title{
\scriptsize{\bf {Mechanics of Solids \\
Vol. 40, No. 6}}\\
\vspace{36pt}
\Large{ELASTOPLASTIC EQUATIONS WITH EXACT CONSISTENCY
OF STRAINS FOR THE ELASTIC STRAIN FRAME AND THE VELOCITIES OF POINTS} }
\author{I.A.Solomeshch,  M.A.Solomeshch}
\date{}
\begin{document}

\maketitle

\begin{abstract}
{\small We obtain an exact strain consistency equation for full, elastic and
plastic strain characteristics that have a clear physical meaning and are
naturally related to stresses. The dynamic equations are represented in a
form that does not use the objective stress rate. This obviates a number of
essential difficulties in the theory of finite elastoplastic strains.}
\bigskip 

\end{abstract}
\section {Introduction}

The present paper gives a detailed justification of the results obtained in
[1]. To characterize them, let us very roughly outline the main
difficulties encountered in mathematical elastoplasticity when passing from
infinitesimal to finite strains and the methods used so far to overcome
these difficulties. A detailed analysis can be found in the critical survey
[2] of the state of the art in the theory by 1990 and in the monograph [3],
which summarizes what has been done in the last fifty years.

Since the above-mentioned difficulties already arise when dealing with
homogeneous isotropic elastic--perfectly plastic materials, we momentarily
restrict ourselves to such materials.

In the Eulerian variables, elastoplastic deformation processes are
described in terms of mass point velocities, material density, selected
measures (characteristics) of full, elastic, and plastic strains, and
stresses. With the advent in 1930 of the Prandtl--Reu\ss\ equation
(see~[4]), the closed system of elastoplasticity equations for
infinitesimal strains was obtained. We single out the following three of
these equations (the remaining three equations remain unchanged under the
transition to finite strains):

A. The strain consistency equation (a tensor equation) relates full,
elastic, and plastic strain characteristics.

B. The stress response function (a tensor equation) expresses the stress
via an elastic strain characteristic.

C. The flow law (a tensor equation) relates the plastic strain to the
stresses.

Under the transition to finite strains, the strain consistency equation
(A) is not true in general ([3], 7.2) and the system is no longer closed.
Moreover, not only measures but also the very notions of elastic and
plastic strain have no precise physical meaning, and there is controversy
(as yet unsettled) as to how exactly they should be introduced~([2],~4A).

In this connection, there has been quite a few general theories (some of
which are presented in [5--11]), a fact that alone shows how unsatisfactory
the situation is. All these theories can be divided into four groups.

{\it 1.} Elastic strain is not considered at all. The plastic strain is
introduced by a constitutive equation of type (C) via a measure that does
not have a precise physical meaning. Two variables (plastic and full
strain) are introduced in the constitutive equation (B) instead of elastic
strain, and the system again becomes closed (e.g., see [5]).

{D. The fact that the strain measure is artificial casts doubt on the
existence (and all the more, the simplicity) of the constitutive equations
(B) and/or (C).}

{\it 2.} The plastic strain is not considered, and so equation (C)
disappears. Instead of (A), using various analogies, one introduces a
constitutive differential equation for the elastic strain, which closes the
system. It is this new equation that is doubtful (e.g., see [6, 7]).

{\it 3.} Both elastic and plastic strains are introduced. One of them is
determined via the other (using the full strain) but has no precise
physical meaning.

In this case, the very definition already gives the strain consistency
equation instead of (A). The system remains closed but has the drawback
described above in (D) (e.g., see [8, 9]).

{\it 4.} For the strain consistency equation (A), one uses the
multiplicative decomposition of the full strain gradient in terms of the
elastic and plastic strain gradients or the related additive decomposition
of the full strain rate in the elastic and plastic strain rates [10].

Serious objections to the first relation can be found in [2], 4A. The
second relation is not true in general and involves the objective stress
rate, resulting in heavy inconsistencies ([2], 4F; [3], 7.2, 7.3).

In the present paper, elastic strain is characterized by the elastic strain
frame $\check a$, which is more informative than the Cauchy--Lagrange tensor. In the
deformed medium, the elastic strain frame field generates the Riemannian
metric of the ``natural, unstressed state." Under sufficiently wide
assumptions, this metric permits one to judge about the natural measure of
deformed material objects (i.e., the measure that they would have after
unloading), without resorting to actual unloading.

We introduce the plastic strain tensor, which has the same physical meaning
in terms of the natural measure as does the elastic strain tensor for
elastic media in terms of the geometric measure.

We obtain the strain consistency equation in frame form, providing an
exact relationship between the full, elastic, and plastic strain
characteristics.

We introduce the relative stress tensor (with respect to the Riemannian
metric generated by an arbitrary nondegenerate frame field), whose
restrictions are the natural stress tensor and the conventional Cauchy
stress tensor and Piola--Kirchhoff second stress tensor.

For media homogeneous and isotropic in the natural state, we suggest a
plastic constitutive law specified by a scalar function $p$ of the
principal stresses and possibly by parameters determined by the strain
history. The presence or absence of yield surfaces is determined by the
properties of~$p$.

Under the assumptions accepted in the paper, we obtain a system of
quasilinear first-order partial differential equations in Eulerian
variables for $\nkg v$ and $\check a$, describing the elastoplastic
deformation process in the entire space--time domain occupied by the
medium.
All equations are given in the form solved for the total time derivative,
which makes the system suitable for numerical solution.\footnote{Note
that the elastic strain frame was introduced for elastic media in
I.~A.~Solomeshch and V.~Sh.~Khalilov, Nonlinear Elasticity Equations in
Strain Frame Components [in Russian], No.~5089-B90, VINITI, Moscow,
1990 and for elastoplastic media in I.~A.~Solomeshch and M.~A.~Solomeshch,
Dynamic Elastoplasticity Equations for the Elastic Strain Frame and the
Velocities of Points [in Russian],  No.~214-B95, VINITI, Moscow,
1995. The latter paper indicates the possible connection of the elastic
strain frame in polycrystal materials with the strain of the crystal
lattice. In the earlier-mentioned papers [6, 7, 9], frames actually
coinciding with the elastic strain frame are introduced under various names
(on the average, the trivector of microstructure variables) to characterize
the elastic strain in such materials.}

\section {Some notation and conventions}

We assume that the real space in which the continuous medium moves is the
three-dimensional Euclidean affine space. We denote it by $E$ and the
associated vector space by $\nkg E$. Let $\langle\cdot,\cdot\rangle$ and $|\cdot|$ be the scalar
product and the corresponding norm in $\nkg E$.

A frame consisting of vectors ${\nkg a}_i\in\,\nkg E$, $i=1,2,3$, will be denoted by
$\check a$ and written as a column matrix, so that
$\check a:=({\nkg a}_1,{\nkg a}_2,{\nkg a}_3)^T$, where $T$ stands for transposition. (For
convenience, the entries of a row matrix are separated by commas.) In the
presence of an ambiguity, we always take a nondegenerate right frame.The
symbols $:=$ and $=:$ stand for " is by definition equal to", where the
object to be defined is written next to the colon.

Let us define the norm of a frame in $E$ by the formula $\|\check a\|:=(|{\nkg a}_1|^2+|{\nkg a}_2|^2+|{\nkg a}_3|^2)^{1/2}$
and the difference of frames by the formula $\check a-\check b:=({\nkg a}_1-{\nkg b}_1,{\nkg a}_2-{\nkg b}_2,{\nkg a}_3-{\nkg b}_3)^T$.
The coefficients (contravariant coordinates) in the expansion of a vector ${\nkg x}\in\,\nkg E$
in the vectors of a frame $\check a$ will be denoted by ${\nkg x}^{\check a}_i$, or, more briefly,
by $x^{\check a}_i$;  $\nkg x^{\check a}:=x^{\check a}:=(x^{\check a}_1,x^{\check a}_2,x^{\check a}_3)^T$.
Thus $\nkg x=x^{\check a}_i \nkg a_i$. Here and in what follows, summation from $1$ to $3$ is assumed over a
repeated nonunderlined index in a monomial.

The scalar product defined on $\nkg E$ can be represented in coordinate form
in any frame $\check{e}$ orthonormal with respect to this product:
$$
  \langle \nkg {\alpha}, \nkg {\beta} \rangle=\alpha^{\check e}_i\beta^{\check e}_i\quad
  \forall \nkg {\alpha} , \nkg {\beta}  \in \nkg E.\eqno(2.1)
$$
Conversely, for an arbitrary nondegenerate frame $\check a$ one can construct a
scalar product (generally different from $\langle\cdot,\cdot\rangle$) in $\nkg E$ by setting
$$
  \langle \nkg {\alpha},\nkg {\beta} \rangle_{\check a}:=\alpha^{\check a}_i\beta^{\check a}_i;\eqno(2.2)
$$
the frame $\check a$ is orthonormal with respect to this scalar product. The
Euclidean affine space obtained by the replacement of the scalar product
$\langle\cdot,\cdot\rangle$ by $\langle\cdot,\cdot\rangle_{\check a}$ will be
denoted by $E_{\check a}$. Obviously, $E_{\check e}=E$ if the frame
$\check e$ is orthonormal in $E$.

The modulus of a vector $\nkg {\alpha} \in \nkg E$ in $E_{\check a}$ is equal to
$$
  |\nkg {\alpha}|_{\check a}:=\langle \nkg {\alpha}, \nkg {\alpha} \rangle_{\check a}^{1/2}.\eqno(2.3)
$$
The volume of the parallelepiped spanned by a frame $\check b$ in the space
$E$ will be denoted by $|\check b|$, and the volume of the same parallelepiped
in $E_{\check a}$ will be denoted by $|\check b|_{\check a}$.

Nondegenerate frames $\check a $ and $\check a^\prime$ generating the same scalar product, i.e., such
that $\langle \nkg {\alpha}, \nkg{\beta}\rangle_{\check a}=\langle \nkg {\alpha}, \nkg {\beta} \rangle_{\check a^\prime}$
for any $\nkg {\alpha},\, \nkg {\beta} \in \nkg E$, will be called {\it equivalent\/}.

For frames $\check a$ and $\check b$, by $\check a^{\check b}$ we denote the matrix whose $j$th
column is $a_j^{\check b}$. Matrices are sometimes indicated by the symbol
$\hat{}\;$. For column matrices $\hat c:=(c_1,c_2,c_3)^T$ and $\hat d:=(d_1,d_2,d_3)^T$
with numerical entries, we write
$$
  (\hat c,\hat d):=c_id_i,\quad
  |\hat c|:=(\hat c,\hat c)^{1/2}\eqno(2.4)
$$

Let a continuous medium occupying an open spatial domain $\mathrm V\subset E$ at
the initial time $t_0$ fill a domain $V_t$ at the final time $t$. The
spatial position of a particle $M$ at time $t_0$ will be denoted by $\mathrm x$
(the Lagrangian variable), and the position of the same particle at time
$t$ will be denoted by $x$ (the Eulerian variable).

The function $x=x(\mathrm x)$ specifying the correspondence between the initial and
final positions of a mass point, as well as the laws of motion considered
in what follows, is assumed to be one-to-one and have an invertible
derivative~$x_\mathrm x^\prime$.

In general, we adhere to the following principle whenever possible: objects
characterizing the medium at the terminal time are denoted by italic Latin
letters, and the corresponding objects at the initial time are denoted by
the same letters typeset in Roman; any function of a particle expressed via
$x$ or $\mathrm x$ is denoted by same letter with the argument indicated where
necessary; notation introduced for some object is later used for other
objects of the same nature. Unless explicitly specified otherwise, we
assume that all functions occurring in the text (including the
derivatives) are continuous, all curves and surfaces are smooth, and all
sets over which integration is carried out are measurable and closed.
Whenever the metric is not indicated explicitly, the metric of the main
space $E$ is meant.

In the following, to reveal elastic and plastic strains of the medium, we
need to compare the measure of material objects in Riemannian spaces that
are different in general. The corresponding mathematical technique is given
in the following section.

\section {The change in the measure of geometric objects under a
mapping of Riemannian affine spaces}

{\it 3.1. The Riemannian metric generated by a frame field.} Let $V$ be an
open domain in an affine space $E$. (Momentarily, it is not important
whether $E$ is Euclidean.) A Riemannian metric on $V$ is usually defined
with the help of a metric form
$$
  g_{ij}(x(\xi))d\xi_id\xi_j\eqno(3.1)
$$
in some curvilinear coordinate system $x=x(\xi)$. Essentially, formula (3.1)
defines a scalar product at any point $x\in V$ (more precisely, on the tangent
space at $x$, i.e., in our case, on $\nkg E$). Therefore, one can define a
Riemannian metric by directly specifying a scalar product at the points of
$V$.

For our purposes, it is convenient to specify each of these products by
specifying a frame orthonormal with respect to this product.

Thus we adopt the following scheme. Let a nondegenerate frame field $\check\alpha(x)$
be given on $V$. For the scalar product at a point $x$ we take the product
$\langle\cdot,\cdot\rangle_{\check\alpha(x)}$, with respect to which $\check\alpha(x)$
is orthonormal (see (2.2)). The
Riemannian space thus defined is said to be generated by the frame field
$\check\alpha(x)$ and will be denoted by $V_{\check\alpha}$. The metric in it will be called the
$\check\alpha$-metric.

The relationship between the coefficients of the metric form and $\check\alpha(x)$ is
as follows:
$$
g_{ij}(x(\xi))=
  \langle \nkg {k}_i(\xi),\nkg {k}_j(\xi)\rangle_{\check \alpha(x(\xi))}
$$
where
$$
\check k(\xi):=\frac{\partial x}{\partial \xi}:=
\left(\frac{\partial x}{\partial \xi_1},\frac{\partial x}{\partial \xi_2},
      \frac{\partial x}{\partial \xi_3}\right)^T.
\eqno(3.2)
$$
Note that $\check k(\xi)$ is the local frame of the curvilinear coordinate system at the
point $x(\xi)$.

The conventional formulas of Riemannian geometry in terms of the frame
$\check\alpha(x)$ acquire the following form. The cosine of the angle between vectors
$\nkg {x}$ and $\nkg {y}$ at a point $x$ is
$$
  \cos(\widehat{\nkg {x},\nkg {y}})_{\check\alpha(x)}:=
  \frac{\langle \nkg {x},\nkg {y}\rangle_{\check\alpha(x)}}{|\nkg { x}|_{\check\alpha(x)}\cdot|\nkg {y}|_{\check\alpha(x)}};\eqno(3.3)
$$
The length of a curve $l$: $x=x(\tau)$, $\tau\in[\tau_0,\tau_1]$, is
$$
  |l|_{\check\alpha}:=\int^{\tau_1}_{\tau_0}|x^\prime(\tau)|_{\check\alpha(x(\tau))}\,d\tau;\eqno(3.4)
$$
The area of a surface $s$: $x=x(u)$, $u=(u_1,u_2)^T\in\sigma\subset \nkg {R}^2$, is
$$
  |s|_{\check\alpha}=\int_\sigma\sqrt{|x^{\prime}_{u_1}|^2_{\check\alpha(x)}\cdot
  |x^\prime_{u_2}|^2_{\check\alpha(x)}-\langle x^\prime_{u_1},x^\prime_{u_2}\rangle^2_{\check\alpha(x)}}\,du;
  \eqno(3.5)
$$
The volume of a domain $v$: $x=x(\xi)$, $\xi=(\xi_1,\xi_2,\xi_3)^T\in\omega\subset \nkg {R}^3$, is
$$
  |v|_{\check\alpha}=
  \int_\omega|\check k(\xi)|_{\check\alpha(x(\xi))}\,d\xi.\eqno(3.6)
$$
If $\check\alpha(x)$ is a constant $\check\alpha_0$, then $V_{\check\alpha_0}$ is the Euclidean
space with scalar product $\langle \nkg {x}, \nkg {y}\rangle_{\check\alpha_0}$.

{\it 3.2. The dilatation coefficients under a change of the Riemannian
metric.} Let $\check\alpha(x)$ and $\check\beta(x)$ be nondegenerate frame fields defined on $V$,
and let $V_{\check\alpha}$ and $V_{\check\beta}$ be the Riemannian spaces generated by these frame
fields. Let us find the dilatation coefficients for the measures of one- to
three-dimensional objects under the transition from their measurement in
the $\check\alpha$-metric to the measurement in the $\check\beta$-metric.

{\it The length dilatation coefficient.} Take an arbitrary vector $\nkg {m}\neq0$.
Let $l$: $x=x(\tau)$, $\tau\in[\tau_0,\tau_1]$ be an arbitrary curve issuing
from the point $x_0$, let $x(\tau_0)=x_0$; $\nkg {m}$ be
the tangent vector to $l$ at $x_0$, and let $l_{\tau^{\ast}}:=x([\tau_0,\tau^{\ast}])$
for each~$\tau^{\ast}\in(\tau_0,\tau_1]$.

The length dilatation coefficient at the point $x_0$ in the direction $\nkg {m}$
is computed with the use of (3.4):
$$
  K_1(x_0,\nkg {m},\check\alpha,\check\beta):=\lim_{\tau^\ast\to\tau_0}
  \frac{|l_{\tau^\ast}|_{\check\beta}}
       {|l_{\tau^\ast}|_{\check\alpha}}=
  \lim_{\tau^\ast\to\tau_0}
  \frac{{\int^{\tau^\ast}_{\tau_0}|x^\prime_\tau|_{\check\beta}d\tau}}
       {{\int^{\tau^\ast}_{\tau_0}|x^\prime_\tau|_{\check\alpha}d\tau}}=
  \frac{|x^\prime_\tau(\tau_0)|_{\check\beta(x_0)}}
       {|x^\prime_\tau(\tau_0)|_{\check\alpha(x_0)}}\,.
$$
The vectors $x^\prime(\tau_0)$ and $\nkg {m}$ are collinear, and hence $x^\prime_\tau(\tau_0)=c \nkg {m}$ for some $c$.
Eventually, we obtain  
$$
  K_1(x_0,\nkg {m},\check\alpha,\check\beta)=
  \frac{|\nkg {m}|_{\check\beta(x_0)}}{|\nkg {m}|_{\check\alpha(x_0)}}.\eqno(3.7)
$$

{\it The volume dilatation coefficient.} Take an arbitrary $\forall x_0\in V$. Let
$v$: $x=x(\xi)$, $\xi\in\omega\subset\bf R^3$, be an arbitrary three-dimensional subdomain of $V$ containing
$x_0$, $x_0=x(\xi_0)$; suppose that $\omega^\ast\subset\omega$ contains $\xi_0$ and
$v_{\omega^\ast}:=x(\omega^\ast)$.

The volume dilatation coefficient at the point $x_0$ is computed with the
use of (3.6):
$$
  K_3(x_0,\check\alpha,\check\beta):=
  \lim_{\omega^\ast\to\xi_0}\frac{|v_{\omega^\ast}|_{\check\beta}}
                                 {|v_{\omega^\ast}|_{\check\alpha}}=
  \lim_{\omega^\ast\to\xi_0}
  \frac{\int_{\omega^\ast}|\check k(\xi)|_{\check\beta(x(\xi))}\,d\xi}
       {\int_{\omega^\ast}|\check k(\xi)|_{\check\alpha(x(\xi))}\,d\xi}=
  \frac{|\check k(x_0)|_{\check\beta(x_0)}}
       {|\check k(x_0)|_{\check\alpha(x_0)}}.
  \eqno(3.8)
$$

The volume in $V_{\check\beta(x_0)}$ of the parallelepiped spanned by $\check k(x_0)$ is equal to
$|\check k(x_0)|_{\check\beta(x_0)}=|\check k^{\check\beta}|$, and similarly,
$|\check k(x_0)|_{\check\alpha(x_0)}=|\check k^{\check\alpha}|$.

We denote $\check\alpha^{\check\beta}=:A$; then $\check\alpha=A^T\check\beta$,
and hence the $\check\alpha$- and $\check\beta$-coordinates of
an arbitrary vector $\nkg {y}$ are related by~$y^{\check\alpha}=A^{-1}y^{\check\beta}$.
Therefore, $|\check k|_{\check\alpha}=|\check k^{\check\alpha}|=
|A^{-1}\check k^{\check\beta}|=|A|^{-1}\cdot|\check k|_{\check\beta}$. Substituting
this into (3.8), we obtain
$$
  K_3(x_0,\check\alpha,\check\beta)=|A|=|\check\alpha(x_0)|_{\check\beta(x_0)}.\eqno(3.9)
$$

{\it The area dilatation coefficient.} Take an arbitrary $\forall x_0\in V$.
Let $s$: $x=x(u)$, $u=(u_1,u_2)\in\sigma\subset{\bf R}^2$, be an arbitrary oriented
surface in $V$ with unit positive $\check\beta$-normal $\nkg {n}$ at the point $x_0$,
$x_0=x(u_0)$; we assume that $\sigma^\ast\subset\sigma$
contains $u_0$ and $s_{\sigma^\ast}:=x(\sigma^\ast)$. The area dilatation coefficient at the point
$x_0$ is computed with the use of (3.5):
$$
\begin{array}{lr}
\qquad K_2(x_0,\nkg {n},\check \alpha,\check\beta):=\lim_{\sigma^{\ast}\rightarrow u_0}
(|s_{\sigma^{\ast}}|_{\check\beta}/|s_{\sigma^{\ast}}|_{\check\alpha})
{=} & 
\vspace{6pt}
\, \\
\qquad =\lim_{\sigma^{\ast}\rightarrow u_0}
\dfrac{\int\limits_{\sigma^{\ast}}
\sqrt{|x^{\prime}_{u_1}|^2_{\check\beta(x)}\cdot
      |x^{\prime}_{u_2}|^2_{\check\beta(x)}-
      \langle x^{\prime}_{u_1},x^{\prime}_{u_2}\rangle^2_{\check\beta(x)}}du}
{\int\limits_{\sigma^{\ast}}
\sqrt{|x^{\prime}_{u_1}|^2_{\check\alpha(x)}\cdot
      |x^{\prime}_{u_2}|^2_{\check\alpha(x)}-
      \langle x^{\prime}_{u_1},x^{\prime}_{u_2}\rangle^2_{\check\alpha(x)}}du}=
& 
\vspace{6pt}
\, \\
\qquad =\left(
\dfrac{\sqrt{|x^{\prime}_{u_1}|^2_{\check\beta(x)}\cdot
      |x^{\prime}_{u_2}|^2_{\check\beta(x)}-
      \langle x^{\prime}_{u_1},x^{\prime}_{u_2}\rangle^2_{\check\beta(x)}}}
      {\sqrt{|x^{\prime}_{u_1}|^2_{\check\alpha(x)}\cdot
      |x^{\prime}_{u_2}|^2_{\check\alpha(x)}-
      \langle x^{\prime}_{u_1},x^{\prime}_{u_2}\rangle^2_{\check\alpha(x)}}}
\right)_{|_{x=x_0}} &\qquad\qquad (3.10)
\end{array}
$$

From this formula, we conclude that the dilatation coefficient is same as
if

1. $\check\beta(x)\equiv\check\beta(x_0)$ and $\check\alpha(x)\equiv\check\alpha(x_0),$,
i.e., both metrics are Euclidean.

2. $s$ lies in the plane $\Pi$ with positive unit $\check\beta$-normal $\nkg {n}$ at the
point $x_0$.

But in this case the integrands in (3.10) are constants and there is no
need to pass to the limit; i.e.,
$$
  K_2(x_0,\nkg {n},\check \alpha,\check \beta)=\frac{|s|_{\check \beta(x_0)}}{|s|_{\check \alpha(x_0)}}\quad
  \forall s\subset\Pi.\eqno(3.11)
$$
Consider the parallelepiped $v$ spanned by the vectors $\nkg {m}$, $\nkg {l}$ (parallel
to $\Pi$), and $\nkg {n}$ forming a nondegenerate right frame at $x_0$. We
denote the parallelogram spanned by $\nkg {m}$ and $\nkg {l}$ by $s$ and the
positive unit $\check\alpha$-normal to $\Pi$ by $\bf n$,
$$
  |v|_{\check \beta}=|s|_{\check \beta} \cdot|\nkg {n}|_{\check \beta}=|s|_{\check \beta},\quad
  |v|_{\check \alpha}=|s|_{\check \alpha}\cdot \langle \nkg n,\prg n\rangle_
{\check \alpha}.
$$

Therefore, in view of (3.11) and (3.9),
$$
  K_2(x_0,\nkg {n},\check \alpha,\check \beta)=
  \frac{|v|_{\check \beta}}{|v|_{\check \alpha}}\cdot\langle \nkg n,\prg n\rangle_{\check \alpha}=
  |\check \alpha|_{\check \beta}\cdot\langle\nkg n,\bf n\rangle_{\check \alpha}.\eqno(3.12)
$$

Let us find $\bf n$. To this end, we transform the equation of the plane $\Pi$
in the $\check\beta$-metric to the equation in the $\check\alpha$-metric.

The equation $\Pi$ in the $\check \beta$-metric has the form
$$
  \langle \nkg {n}, \nkg {x}\rangle_{\check\beta}=0,\eqno(3.13)
$$
where $x-x_0=:\nkg {x}$, $\langle \nkg {n}, \nkg {x}\rangle_{\check\beta}=(n^{\check\beta},x^{\check\beta})$. \
$x^{\check\alpha}=A^{-1}x^{\check\beta}$, where $A=\check\alpha^{\check\beta}$.
Therefore, $\langle \nkg {n}, \nkg {x}\rangle_{\check\beta}=
(n^{\check\beta},Ax^{\check\alpha})=(A^Tn^{\check\beta},x^{\check\alpha})$.
If we now introduce the vector $\nkg {N}$ with $\check\alpha$-coordinates
$$
  N^{\check\alpha}:=A^Tn^{\check\beta},\eqno(3.14)
$$
then $\langle \nkg {n}, \nkg {x} \rangle_{\check\beta}=(N^{\check\alpha},x^{\check\alpha})=
\langle \nkg {N}, \nkg {x}\rangle_{\check\alpha}$.

By substituting this into (3.13) and by normalizing $\nkg {N}$ in the $\check\alpha$-norm,
we obtain the equation of $\Pi$ in the $\check\alpha$-metric:
$$
 \langle \nkg {N}/|N^{\check \alpha}|, \nkg {x} \rangle_{\check\alpha}=0 .
$$
Hence $\prg n=\nkg {N}/|N^{\check \alpha}|$, and using (3.14), we obtain
$$
  \mathrm n^{\check\alpha}=\frac{A^Tn^{\check\beta}}{|A^Tn^{\check\beta}|}.\eqno(3.15)
$$

Now, using the expression of the $\check\alpha$-coordinates via the
$\check\beta$-coordinates and (3.15), we have
\begin{eqnarray*}
\langle{\nkg {n}},{\prg n}\rangle_{\check \alpha}=
(n^{\check \alpha},\mathrm n^{\check \alpha})=
(A^{-1}n^{\check \beta},A^Tn^{\check\beta})/|A^Tn^{\check\beta}|\\
=|\nkg {n}|^2_{\check\beta}/|A^Tn^{\check\beta}|=1/|A^Tn^{\check\beta}|.
\end{eqnarray*}
By substituting this into (3.12), we obtain
$$
  K_2(x_0, \nkg {n},\check\alpha,\check\beta)=
  \frac{|\check\alpha(x_0)|_{\check\beta(x_0)}}{|A^Tn^{\check\beta}|},\eqno(3.16)
$$
where $A=\check\alpha(x_0)^{\check\beta(x_0)}$.

\section {The strain frame. The relative stresses}

{\it 4.1.} Material objects (curves, surfaces, three-dimensional regions)
are deformed with respect to the initial state as a result of the
displacement $x=x(\mathrm x)$. One can characterize the resulting strain by comparing
the measures (lengths, areas, volumes) of these objects, as well as the
angles between material curves, at time $t$ and the initial time $t_0$.
Therefore, we characterize all above-mentioned objects in the strained
state, i.e., at time $t$, both by their geometric measure (length, area,
etc.) at time $t$ and by their initial measure at time $t_0$.

Clearly, the specification of $x(\mathrm x)$ or $x^\prime(\mathrm x)$ completely determines the
terminal strain of the medium. In turn, the operator $x^\prime(\mathrm x)=:\mathcal A(\mathrm x)$ at each point is
determined by the specification of an arbitrary nondegenerate frame
~$\check\mathrm a(\mathrm x)$ and the frame
$$
  \check{a}(x):=\left(\mathcal A(\mathrm{x}){\bf a}_1(\mathrm{x}),
\mathcal A(\mathrm{x}){\bf a}_2(\mathrm{x}),
\mathcal A(\mathrm{x}){\bf a}_3(\mathrm{x})\right)^T
=:\mathcal A(\mathrm{x})\check{\mathrm{a}}(\mathrm{x});\eqno(4.1)
$$
one even does not need to require that the function $\check\mathrm a(\mathrm x)$ is continuous.

However, if the frames $\check\mathrm a(\mathrm x)$ are orthonormal, then, to specify the strain
of the medium at time $t$, it suffices to define the frame $\check a(x)$ alone at
each point $x\in V_t$. (Note that $\check a(x)=\check a(\mathrm x)$ for $x=x(\mathrm x)$;
i.e., $\check a(x)$ and $\check a(\mathrm x)$ specify the same function of a material point
expressed via the Eulerian and Lagrangian coordinates, respectively.)

{\it 4.2.} Let us prove this. The computation of the geometric measure of
the deformed (i.e., given at time $t$) material objects does not encounter
difficulties. Therefore, assuming that the frame field $\check a(x)$ is given, we focus
our attention on the computation of their initial measure.

Let $\mathrm l$, $\mathrm l_i$ ($i=1,2$), $\mathrm s$, and $\mathrm v$ be oriented material curves, a surface, and a
three-dimensional region given at the initial time in $\mathrm V$; let $l:=x(\mathrm l)$,
$l_i:=x(\mathrm l_i)$, $s:=x(\mathrm s)$, and $v:=x(\mathrm v)$ be the same objects at the terminal time; we
suppose that $\mathrm l_1$ and $\mathrm l_2$ meet at the point $\mathrm x_0$; finally, $x_0=x(\mathrm x_0)$;
$\mathrm x=\mathrm x(\tau)$; $\mathrm x=\mathrm x_i(\tau)$, $\tau\in[\tau_1,\tau_2]$;
$\mathrm x=\mathrm x(u)$, $u(u_1,u_2)\in\sigma\subset{\bf R}^2$, and $\mathrm x=\mathrm x(\xi)$,
$\xi(\xi_1,\xi_2,\xi_3)\in\omega\subset{\bf R}^3$ are parametrizations of $\mathrm l$, $\mathrm l_i$, $\mathrm s$, and $\mathrm v$.

To denote the geometric measure of (length, area, or volume) of
1--3-dimensional objects, we shall use vertical bars. The initial measure
of such objects given in the strained state will be denoted in the same way
but with the subscript $t$. For example, $|\mathrm l|$ is the geometric length of
$\mathrm l$, $|v|$ is the geometric volume of $v$, and $|s|_t$ is the geometric
area of $s$ (i.e., $|\mathrm s|$). An geometric angle between vectors $\nkg {n}$ and
$\nkg m$ will be denoted by $(\widehat{\nkg n,\nkg m})$, and the initial angle between the same
vectors issuing from a point $x$ will be denoted by $(\widehat{\nkg n,\nkg m})_{x,t}$. The adjective
"geometric" for a measure will usually be omitted.

First, let us compute the initial angle between the curves $l_i$ at the
point $x_0$. For the parametrization of $l_i$ one can take $x=x(\mathrm x_i(\tau))=:x_i(\tau)$.
Let $\tau_0$ be the parameter value corresponding to $\mathrm x_0$ on the curves $\mathrm l_i$; then it
also corresponds to $x_0$ on $l_i$.

The derivatives $x_i^\prime(\tau_0)=:\nkg m_i$ are the positive tangent vectors to the curves $l_i$
at the point $x_0$, and $\mathrm x_i^\prime(\tau_0)=:{\bf m}_i$ is the tangent vector to $l_i$ at the
point $\mathrm x_0$; here $\nkg m_i$ and $\prg m_i$ are related by
$$
  \nkg m_i=x^\prime_i(\tau_0)=x^\prime_\mathrm x(\mathrm x_0)
  \mathrm x^\prime_i(\tau_0)=\mathcal A(\mathrm x_0)\prg m_i.\eqno(4.2)
$$

We need the following lemma.

{\it Lemma. If $\mathcal A$ is a linear invertible operator on $\nkg E$, $\check\mathrm a$ is a
nondegenerate frame, and $\check a=\mathcal A\check\mathrm a$, then for any $\nkg {\alpha}$, $ \nkg {\beta} \in \nkg E$
and an arbitrary frame $\check b$ one has
$$
\qquad \alpha^{\check{\mathrm{a}}}=
  (\mathcal A \nkg {\alpha})^{\check{a}}\eqno(4.3)
$$
$$
<\nkg {\alpha},\nkg {\beta}>_{\check{\mathrm a}}
     =<\mathcal A\nkg {\alpha},\mathcal A\nkg {\beta}>_{\check{a}}, \;\;
|\nkg {\alpha}|_{\check{\mathrm a}}=|\mathcal A\nkg {\alpha}|_{\check a} \eqno(4.4)
$$
$$ |\check{b}|_{\check{\mathrm a}}=|\mathcal A\check{b}|_{\check{a}}
\eqno(4.5)
$$}
{\it Proof.} We have $\nkg {\alpha}=\alpha_i^{\check\mathrm a}{\bf a}_i$. Therefore,
$\mathcal A \nkg {\alpha}=\alpha_i^{\check\mathrm a}{\mathcal A}{\bf a}_i=\alpha_i^{\check\mathrm a}\nkg a_i$;
i.e., $(\mathcal A \nkg {\alpha})^{\check a}=\alpha^{\check\mathrm a}$. Assertions (4.4)
readily follow from (4.3) by virtue of definitions~(2.2) and~(2.3).
Further, using the Gram determinant, we have
$$
  |\check b|_{\check\mathrm a}=|\,\langle \nkg {b}_i,\nkg {b}_j\rangle_{\check\mathrm a}|^{1/2}=
  |\,\langle\mathcal A \nkg {b}_i,\mathcal A \nkg {b}_j\rangle_{\check a}|^{1/2}=|\mathcal A\check b|_{\check a},
$$
and the proof of the lemma is complete.

Using (2.1)--(2.3), we compute the cosine of the initial angle between the
curves $l_i$ at the point $x_0$:
$$
  \cos(\widehat{\nkg m_1,\nkg m}_2)_{x_0,t}=\cos(\widehat{{\bf m}_1,{\bf m}}_2)=
  \frac{\langle {\bf m}_1,{\bf m}_2\rangle}{|{\bf m}_1|\cdot|{\bf m}_2|}=
  \frac{\langle {\bf m}_1,{\bf m}_2\rangle_{\check\mathrm a(\mathrm x_0)}}
  {|{\bf m}_1|_{\check\mathrm a(\mathrm x_0)}\cdot|{\bf m}_2|_{\check\mathrm a(\mathrm x_0)}}.
$$
Thus, using (4.4) and (4.2), we obtain
$$
  \cos(\widehat{\nkg m_1,\nkg m}_2)_{x_0,t}=
  \frac{\langle \nkg {m}_1,\nkg m_2\rangle_{\check a(x_0)}}
  {{|\nkg m_1|_{\check a(x_0)}}\cdot|\nkg m_2|_{\check a(x_0)}}.\eqno(4.6)
$$
It is easy to express the initial length of the curve $l$ via its
equation
$$
  |l|_t=|\mathrm l|=\int_{\tau_1}^{\tau_2}|\mathrm x^\prime(\tau)|\,d\tau=
  \int_{\tau_1}^{\tau_2}|\mathrm x^\prime(\tau)|_{\check{\mathrm a}(\mathrm x(\tau))}\,d\tau.
$$

Since $x^\prime(\tau)=x_\mathrm x^\prime(\mathrm x(\tau))\mathrm x^\prime(\tau)=
\mathcal A(x(\tau))\mathrm x^\prime(\tau)$, we obtain, in view of (4.4),
$$
  |l|_t=\int_{\tau_1}^{\tau_2}|x^\prime(\tau)|_{\check a(x(\tau))}\,d\tau.\eqno(4.7)
$$
In a similar way, we find the expressions for the initial area and volume:
$$
 |s|_t=|\mathrm s|=\int_\sigma\sqrt{\langle\mathrm x_{u_1}^\prime,\mathrm x_{u_1}^\prime\rangle\cdot
    \langle\mathrm x_{u_2}^\prime,\mathrm x_{u_2}^\prime\rangle-\langle\mathrm x_{u_1}^\prime,\mathrm x_{u_2}^\prime\rangle^2}\,du.$$
Let us replace the scalar product $\langle\cdot,\cdot\rangle$ by
$\langle\cdot,\cdot\rangle_{\check\mathrm a(\mathrm x)}$. (Recall that the frames
$\check\mathrm a(\mathrm x)$ are orthonormal.) Then we apply~(4.4) and take into account the fact
that $x_{u_i}^\prime=x_\mathrm x^\prime\,\mathrm x_{u_i}^\prime=\mathcal A\mathrm x_{u_i}^\prime$ to obtain
$$
  |s|_t=\int_\sigma\sqrt{\langle x_{u_1}^\prime,x_{u_1}^\prime\rangle_{\check a(x)}\cdot
    \langle x_{u_2}^\prime,x_{u_2}^\prime\rangle_{\check a(x)}-\langle x_{u_1}^\prime,
    x_{u_2}^\prime\rangle_{\check a(x)}^2}\,du.\eqno(4.8)
$$
The coordinate frame of the curvilinear coordinates $\mathrm x=\mathrm x(\xi)$ at the point
$\mathrm{\check k (x)}:=\left (\pd {\mathrm x} {\xi_1},\pd {\mathrm x}
{\xi_2},\pd {\mathrm x} {\xi_3} \right )^T=:\pd {\mathrm x} {\xi},\;\;$; therefore, $|v|_t=|\mathrm v|=\int_\omega|\check\mathrm k(\mathrm x)|\,d\xi$.
Since the frames $\check\mathrm a(\mathrm x)$ are orthonormal,
it follows that $E=E_{\check\mathrm a(\mathrm x)}$, and, using (4.5), we obtain
$|\check\mathrm k(\mathrm x)|=|\check\mathrm k(\mathrm x)|_{\check\mathrm a(\mathrm x)}
=|\mathcal A(\mathrm x){\check\mathrm k(\mathrm x)}|_{\check a(x)}$. Since
$$
  \mathcal A(\mathrm x)\mathrm{\check k (x)}=x_{\mathrm x}^\prime (\mathrm x)
                          \pd {\mathrm x} {\xi}=
\left (x_{\mathrm x}^\prime (\mathrm x)\mathrm{x}_{\xi_1}^\prime,
       x_{\mathrm x}^\prime (\mathrm x)\mathrm{x}_{\xi_2}^\prime,
       x_{\mathrm x}^\prime (\mathrm x)\mathrm{x}_{\xi_3}^\prime \right )^T=:
\check{k}(x)
$$
is a coordinate frame of the curvilinear coordinates $x=x(\xi)$ at the point
$x$, we have
$$
  |v|_t=\int_\omega|\check k(x)|_{\check a(x)}\,d\xi.\eqno(4.9)  
$$

Formulas (4.6)--(4.9) show that, given the frame field $\check a(x)$, one can
compute the initial measure of 1--3-dimensional material objects and the
initial angle between the material curves from the equations of these
objects in the strained state (i.e., at time $t$). It follows that by
specifying the field $\check a(x)$ one completely determines the strain of the
medium at time $t$.

On the other hand, Eqs.~(4.6)--(4.9) show (see Section 3.1) that the
original metric for deformed material objects is the metric generated by
the frame field $\check a(x)$, i.e., the metric obtained in the Riemannian affine
space $V_{t,\check a}$ by the replacement of the scalar product $\langle\cdot,\cdot\rangle$
common for all points in $V_t$ by the field of scalar products $\langle\cdot,\cdot\rangle_{\check a(x)}$,
individual for each point $x\in V_t$.

{\it 4.3.} We take a point $x\in V_t$ and a vector $\nkg m\neq0$. Let $l$ be a material
curve at time $t$ for which $\nkg m$ is the tangent vector at the point $x$.
We consider the length dilatation coefficient for this curve at time $t$,
equal to the ratio $|l|/|l|_t$ of its length at time $t$ to its length in the
initial state.

It turn out (see Section 3.2) that the limit of this coefficient as the
curve shrinks along itself to the point $x$, is independent of the choice
of $l$ and depends only on $x$ and $\nkg m$. We refer to this limit as the
length dilatation coefficient at the point $x$ at time $t$ (i.e., at the
point $(x,t)$) in the direction $\nkg m$ and denote it by $K_1(x,\nkg m)$.

The area dilatation coefficient $K_2(x,\nkg n)$ at a point $(x,t)$ of a material
surface with unit normal $\nkg n$ at the point $x$ at time $t$ and the volume
dilatation coefficient $K_3(x)$ at the point $(x,t)$ are introduced in a similar
way.

Since for each orthonormal frame $\check e$ the scalar product
$\langle\cdot,\cdot\rangle$ coincides with $\langle\cdot,\cdot\rangle_{\check e}$,
it follows that the geometric measure on $V_t$ coincides with the measure
on $V_{t,\check e}$ for any orthonormal frame field $\check e(x)$ on $V_t$.
Furthermore, the initial metric coincides
with the metric in $V_{t,\check a}$. Therefore, $K_i$ is the measure dilatation
coefficient for the passage from the $\check a$-metric to the $\check e$-metric
(i.e., the  geometric metric) in $V_t$. Hence, according to formulas (3.7),
(3.16), and (3.9) with $\check \alpha=\check a$ and $\check\beta=\check e$, we have
$$
  K_1(x,\nkg m)=\frac{|\nkg m|}{|\nkg m|_{\check a}},\quad
  K_2(x,\nkg n)=\frac{|\check a|}{|A^T n^{\check e}|},\quad
  K_3(x)=|\check a|,\eqno(4.10)
$$
where $A=\check a^{\check e}$.

Formula (4.6), together with (4.10), shows that $\check a(x)$ completely determines
the strain of the medium at the point~$(x,t)$.

Therefore, $\check a(x)$ will be called the strain frame. Let us give a complete
definition taking into account (4.1) and using derivatives with respect to
a vector.

The strain frame of the medium at a point $x$ at time $t$ is defined as
$$
 \begin{array}{l}
\check a(x):=\dfrac{\partial x(\mathrm x)}{\partial \check{\mathrm a}}:=
\left(\frac{\partial x(\mathrm x)}{\partial {\bf a}_1},\,
      \frac{\partial x(\mathrm x)}{\partial {\bf a}_2},\,
      \frac{\partial x(\mathrm x)}{\partial {\bf a}_3}\right)^T=
\left(x^\prime (\mathrm x){\bf a}_1,\,
      x^\prime (\mathrm x){\bf a}_2,\,
      x^\prime (\mathrm x){\bf a}_3\right)^T \\=
x^\prime (\mathrm x)\check{\mathrm a}\, ,
\end{array}\eqno(4.11)
$$
where $\check{\mathrm a}$ is an arbitrary orthonormal frame.

The physical meaning of the strain frame $\check a$ is clear from the very definition
of the derivative with respect to a vector in conjunction with (4.10)
(which is in generally not necessary) for $\nkg m=\nkg a_i$.

The vectors of the frame $\check a$ are the tangents at the point $x$ to three
strained material fibrils issuing from $x$ and mutually orthogonal at the
initial time. Moreover, $|\nkg a_i|$ is the dilatation coefficient of the material
at $x$ at time $t$ in the direction $\nkg a_i$.

Restating the last paragraph in Section 4.2, we can say that the
specification of the strain frame field permits one to introduce a
Riemannian metric on $V_t$ such that the measurement of angles between
deformed material curves and of lengths, areas, and volumes of deformed
1--3-dimensional material objects in this metric gives the values of the
respective characteristics for the objects in question prior to strain
without resorting to their initial state.

The strain frame is not unique. By taking another orthonormal frame $\check\mathrm a^\prime$
in the undeformed medium, we obtain another strain frame
$\check a^\prime:=\mathcal A\check\mathrm a^\prime$ by (4.11).
Obviously, the frames $\check a$ and $\check a^\prime$ characterize the same strain.

Note that the values of the expressions (4.6) and (4.10) do not change if
the frame $\check a$ is replaced by an equivalent frame.

{\it 4.4.} We take a point $x$ in the deformed medium and a unit vector
$\nkg n$. The stress per unit geometric area at the point~$x$ at time $t$ on
the oriented plane with positive normal $\nkg n$ will be called the geometric
stress (the Cauchy stress~[12], or the true stress [13]) and will be
denoted by $\nkg {\sigma}_{\slin {\nkg n}}(x)$. The stress per unit initial area will be referred to as
relative (to the initial area).

The geometric measure on $V_t$ coincides with the measure on $V_{t,\check e}$ for any
orthonormal frame field $\check e(x)$ on $V_t$, and the initial measure, as was
already mentioned, coincides with the measure on $V_{t,\check a}$.

Generalizing, for an arbitrary nondegenerate frame field $\check b(x)$ defined on
$V_t$, we introduce the stress $\nkg {u}_{\slin {\nkg n}}(x,\check b)$ per unit
area in the Riemannian space $V_{t,\check b}$, which will be called the relative
stress (relative to $\check b(x)$). The symbol $\check b$ is omitted if it is clear
what frame is meant.

The geometric stress and the stress relative to the initial area are the 
special cases of $\nkg {u}_{\slin {\nkg n}}(x,\check b)$ for $\check b(x)=\check e(x)$
and $\check b(x)=\check a(x)$, respectively.

{\it 4.5.} We introduce the following notation. Let $\nkg b_i$ be one
of the vectors in the frame $\check b$. By $\nkg n_i(\check b)$ we denote the unit normal
to the plane spanned by the other two vectors of $\check{b}$; we assume that
$\nkg n_i(\check b)$ points to the same side of this plane as $\nkg b_i$.
Next, $\nkg u_i(x,\check b):=\nkg u_{\slin {\nkg n_i(\check b)}}(x,\check b)$ are the relative stresses
on the faces of the frame $\check b(x)$ at
the point $(x,t)$; $\nkg n$ is the unit positive normal to a plane $\Pi$ passing
through $x$; $\bf n$ is the unit positive normal to the same plane in the
metric $V_{t,\check b}$ (it will be called the $\check b$-normal);
$(\mathrm n_1,\mathrm n_2,\mathrm n_3)^T:={\bf n}^{\check b}$.

One can prove the generalized Cauchy relation
$$
  \nkg {u}_ {\slin {\nkg n}}(x,\check b)=\nkg {u}_i(x,\check b)\mathrm n_i\eqno(4.12)
$$
for the relative stresses in completely the same way as the classical
Cauchy relation between the Cauchy (geometric) stresses on the plane $\Pi$
and the faces of $\check e$ for an orthonormal frame $\check e$.

Introducing the frame
$$
  \check u(x,\check b):=\left(\nkg u_1(x,\check b),\,\nkg u_2(x,\check b),\,\nkg u_3(x,\check b)\right)^T\eqno(4.13)
$$
of relative stresses on the faces of $\check b(x)$, we can rewrite the dependence
(4.12) in the form
$$
  \nkg {u}_ {\slin {\nkg n}}(x,\check b)=\check u(x,\check b)^T{\prg n}^{\check b}.\eqno(4.14)
$$

{\it 4.6.} Let us establish the connection between the relative stress
frames for different initial frame fields. We take a point $x\in V_t$ and two
arbitrary nondegenerate frames $\check a$ and $\check b$ at $x$. (The values of the
frames at any other points are irrelevant.) Suppose that $\check u(x,\check a)$ is known.
Let us compute the stress $\nkg u_i(x,\check b)$ on the $i$th face of $\check b$. We denote the
unit positive normal, the $\check a$-normal, and the $\check b$-normal to this face
by $\nkg m$, $\prg n$, and $\nkg n$, respectively; next, $A:=\check a^{\check b}$,
$B:=\check b^{\check a}$, and~$\hat i$ is
the column with unit in the $i$th position and zeros in the remaining
positions. Then $\nkg n=\nkg b_i$ and hence
$$
  n^{\check b}=\hat i;\eqno(4.15)
$$
since $\check a=A^T\check b$, it follows that $\check b=A^{-T}\check a$ and hence
$$
  B=A^{-1}.\eqno(4.16)
$$

By formulas (3.15) and (3.16), the area dilatation coefficient for the
transition from the $\check a$-measure to the $\check b$-measure is equal to
$$
  K_2(x,\nkg n)=\frac{|\check a|_{\check b}}{|A^Tn^{\check b}|}\quad\hbox{and}\quad
  \mathrm n^{\check a}=\frac{A^Tn^{\check b}}{|A^Tn^{\check b}|}.\eqno{(4.17)}
$$
Using (4.17), (4.14), and (4.15), we obtain
\begin{eqnarray}
\nkg u_i(x,\check b)=K_2^{-1}(x,\nkg n)\nkg u_{\slin {\nkg m}}(x,\check a)
{=}
\frac{|A^Tn^{\check b}|}{|\check a|_{\check b}}
\check u(x,\check a)^T \mathrm n^{\check a}
{=}\nonumber\\
=|\check a|_{\check b}^{-1}\check u(x,\check a)^TA^T\hat i=
|\check a|_{\check b}^{-1}\check u(x,\check a)^T(a_{i1},a_{i2},a_{i3})^T=
|\check a|_{\check b}^{-1}\nkg u_j(x,\check a)a_{ij}\nonumber
\end{eqnarray}and since, according to (4.16), the $\check b$-volume $|\check a|_{\check b}$ of the
parallelepiped spanned by $\check a$ is equal to the determinant~$|A|=|B|^{-1}$, we have
$$
  \check u(x,\check b)=|B|B^{-1}\check u(x,\check a).\eqno(4.18)
$$
This is the desired relation between the relative stresses.

Let us clarify the relationship between the matrice $U_{\check{a}}$ of decomposition of
$\check u(x,\check a)$ with respect to the frame $\check a$
and matrice $U_{\check{b}}$ of decomposition of $\check u(x,\check b)$ with respect to the frame $\check b$.
Successively substituting $\check u(x,\check a)=U_{\check a}^T \check a$ and
$\check a=B^{-T}\check b$ into (4.18), we have
$\check u(x,\check b)=|B|B^{-1}U_{\check a}^T B^{-T}\check b$ and hence
$U_{\check b}=|B|B^{-1}U_{\check a} B^{-T}$. Thus the matrix of decomposition of the frame of
relative stresses on the faces of the frame $\check a$ with respect to $\check a$
changes under the replacement of $\check a$ by $\check b$ as the matrix of some
contravariant pseudotensor of rank~$2$ and weight~$1$. It will be called the
relative stress pseudotensor at the point~$x$.

\section {The equilibrium equations with respect to \\ an arbitrary
frame field}

Suppose that a nondegenerate frame field $\check b(x)$ of class $C^1$ is given in
$V$ and an orthonormal frame $\check e$ is fixed. Let $\nkg F(x)$ and $ \nkg {\Psi}(x,\check b)$ be the
mass force density at the point $x$ per unit volume (geometric density) and
per unit relative~$\check b$-volume (relative density), respectively, and let
$\nkg {\sigma}_i(x):=\nkg u_i(x,\check e)$ be the Cauchy (geometric) stress at $x$ on the
$i$th face of $\check e$.

Since the volume dilatation coefficient (3.9) for the transition from the
$\check b$-volume to the geometric volume is equal to $|\check b|$, we have
$\nkg {\Psi}(x,\check b)=\nkg F(x)|\check b|$.

The force and moment equilibrium equations for geometric stresses (e.g.,
see [13]) read
$$
  \pd{\nkg {\sigma}_i}{x_i}+\nkg F(x)=0,\quad   
  \Sigma=\Sigma^T,\eqno(5.1)
$$
where $\Sigma:=\check\sigma^{\check e}$ and the $x_i$ are the coordinates of $x$ in the Cartesian
coordinate system with coordinate frame $\check e$. By~(4.18), we have
$$
  \check\sigma=|\check b\,|^{-1}B\check u,\eqno(5.2)
$$
where $B=\check b^{\check e}$ and hence
$\nkg {\sigma}_i(x)=|\check b(x)|^{-1}b_{ij}(x)\nkg u_j(x,\check b)$. Therefore,
$$
  \pd{\nkg {\sigma}_i}{x_i}=\pd{\nkg {\sigma}_i}{\nkg e_i}
  =|\check b\,|^{-1}\pd{b_{ij}}{\nkg e_i}\nkg u_j+b_{ij}\pd{(|\check b|^{-1}\nkg u_j)}{\nkg e_i}.\eqno(5.3)
$$
The first term in (5.3) is equal to
$$
  |\check b|^{-1}\nkg u_j\,div\, \nkg b_j.\eqno(5.4)
$$
Let $\delta_{ij}$ be the Kronecker delta, and let $b^{ij}$ be the entries of the
matrix $B^{-1}$. Using the linearity of the derivative with respect to a
vector in the vector and the fact that $\check b=B^T\check e$, i.e.,
$$
  \check e=B^{-T}\check b\eqno(5.5)
$$
and hence $\nkg e_i=b^{ki}(x)\nkg b_k(x)$, we see that the second term in (5.3) is equal to
$$
  b_{ij}b^{ki}\pd{(|\check b\,|^{-1}\nkg u_j)}{\nkg b_k}=
  \delta_{jk}\left(\pd{\nkg u_j}{\nkg b_k}|\check b\,|^{-1}-|\check b\,|^{-2}\pd{|\check b\,|}{\nkg b_k}\nkg u_j\right)=
  |\check b\,|^{-1}\left(\pd{\nkg u_j}{\nkg b_j}-\pd{\ln|\check b\,|}{\nkg b_j}\nkg u_j\right).
$$
By substituting (5.4) into (5.3) and the resulting relation into the first
equation in (5.1), we obtain the force equilibrium equation for the
relative stresses:    
$$
  \pd{\nkg u_j}{\nkg b_j}+\left(\,div\, \nkg b_j-\pd{\ln|\check b|}{\nkg b_j}\right)\nkg u_j+\nkg {\Psi}=0.\eqno(5.6)
$$

Let us compute the matrix $\check u^{\check b}=:U$. By substituting the expression (5.5) into
the equation $\check\sigma=\Sigma^T\check e$ and then the resulting relation into (5.2), we have
$\check u=|\check b|B^{-1}\Sigma^TB^{-T}\check b$. We see that the matrix $U$ is equal
to $|\check b|B^{-1}\Sigma B^{-T}$ and hence (see (5.1)) is symmetric.

Therefore, the moment equilibrium equation for the relative stresses is
also reduced to the symmetry condition for a matrix, in this case, for the
matrix $U$:
$$
  U=U^T.\eqno{(5.7)}
$$

{\it Remark.} The arbitrary nondegenerate frame field $\check b(x)$ occurring in
Eq.~(5.6) need not be the coordinate frame field of any curvilinear
coordinate system.

\section {The elastic strain frame. The elastic state equation}

{\it 6.1.} We momentarily assume that the medium in question is elastic,  
was in the natural (unstressed) state at the initial time, and was
subjected only to elastic strains until time $t$. In this case, the initial
metric of the deformed medium will be referred to as the natural metric;
accordingly, the stress $\nkg u_ {\slin {\nkg n}} (x,\check a)$ per unit natural area will be called the
natural stress (it is also known as the conditional stress vector [13] and
the first Piola--Kirchhoff stress vector [12]) and will be denoted by
$\nkg t_ {\slin {\nkg n}}(x)$. The density $\nkg {\Psi}(x,\check a)$ of mass forces per unit natural volume will also
be called the natural density and will be denoted by $\nkg {\Phi}(x)$.

Now the strain frame defined by formula (4.11) completely characterizes the
elastic strain of the medium at the point $(x,t)$. It will be called the
elastic strain frame.

Since only natural stresses and densities are used in what follows, we
usually omit the word ``natural."

The Cauchy relation (4.12) passes for the natural stresses into the
well-known relation [12, 13]
$$
  \nkg t_ {\slin {\nkg n}}(x_0)=\nkg t_i(x_0,\check a)\mathrm n_i,\eqno(6.1)    
$$
where $\nkg t_i(x_0,\check a):=\nkg t_{\slin {\nkg n_i(\check a)}}(x_0)$, $\bf n$ is the unit
$\check a$-normal at the point $x$ to the plane
orthogonal to $\nkg n$, and $(\mathrm n_1,\mathrm n_2,\mathrm n_3)^T:=\mathrm n^{\check a}$.

Recall that, according to (4.17),
$$
  \mathrm n^{\check a}=\frac{A^Tn^{\check e}}{|A^Tn^{\check e}|},\quad
  A=\check a^{\check e}\eqno(6.2)
$$
for $\check b=\check e$.

We assume that the medium in the natural state is homogeneous and isotropic
and that the stress at an arbitrary point $x$ of the deformed medium is
completely determined by strain at $x$, i.e., by the elastic strain frame
$\check a(x)$.

Using the frame $\check t(x,\check a)$ of natural stresses at the point $x$ on the faces of
the frame $\check a(x)$, we introduce the stress matrix
$\hat t(x,\check a):=(t_{ij}(x,\check a)):=\check t^{\check a}(x,\check a)$. Under our
assumptions about the medium, the strain frame $\check a(x)$ determines the stress matrix
by the following chain of relations [12]:
$\check a\to\hat\varepsilon^e(\check a):=(\varepsilon^e_{ij}(\check a)):=
(\tfrac12(\langle\nkg a_i, \nkg a_j\rangle-\delta_{ij}))$ (the Cauchy--Lagrange
elastic strain matrix);
$\hat\varepsilon^e\to\hat f(\hat\varepsilon^e):=(f_{ij}(\hat\varepsilon^e))$,
where $f_{ij}$ are functions of class $C^1$
characterizing the elastic properties of the medium;
$\hat t(x,\check a(x))=\hat f(\hat\varepsilon^e(\check a(x)))$.

Thus the equations of the elastic state of the medium can be represented in
the form
$$
  \hat t(x,\check a(x))=\hat f(\hat\varepsilon^e(\check a(x))),\eqno(6.3)
$$
or
$$
  \nkg t_i(x,\check a(x))=f_{ji}(\hat\varepsilon^e(\check a(x)))\nkg a_j(x),\quad
  i=1,2,3.\eqno(6.4)
$$

Thus the matrix $\hat t(x,\check a)$, which is a special case of $U$ (see Section 5), is
symmetric, and the principal stress frame is coaxial with the principal
strain frame.

Since in our case the functions $\hat t$, $\check t$, and $\nkg t_i$ depend only on the
frame $\check a(x)$, the first argument $x$ of these functions can be omitted in
what follows.

{\it 6.2.} Now consider an elastoplastic medium in the process of
elastoplastic deformation. First, we assume  that, at any time $t$ of the
process, the medium, or at least a sufficiently small neighborhood of any
point of the medium, can be unloaded to a natural state by an elastic
deformation. We treat this natural state and the state at time $t$ as the
initial undeformed state and the state obtained from it by an elastic
deformation, respectively. Then (see 6.1) there exists an elastic strain
frame at time $t$ at each point of the medium.

There also exists a similar frame field under more general assumptions. In
general, even a small neighborhood of a point cannot be unloaded to an
unstressed state. There always remain residual stresses. However, one can
assume that these stresses tend to zero as the neighborhood shrinks into
the point, and so the unloaded states tend to an unstressed state.

We take an arbitrary point $x$ of the deformed medium at time $t$. Let $v$
be a neighborhood of $x$. Let us unload the part of the medium contained in
$v$ and consider the strain frame $\check a_v(x,t)$ at $x$, corresponding to the
displacement of the unloaded part to its position at time $t$. Suppose that
there exists a limit $\lim\check a_v(x,t)=:\check a(x,t)$ as $v\to x$ and that
the field $\check a(x,t)$ is continuous.

Now consider an arbitrary material curve in the deformed medium. We cover
it by finitely many balls of radius $r$ and cut into finitely many partial
curves each of which entirely lies in one of the balls. By unloading the
part of the medium lying in a ball, we obtain the "unloaded" length of the
corresponding partial curve. We define the "unloaded" length of the entire
curve as the sum of unloaded lengths of its parts.

One can show that the unloaded lengths of the curve tend as $r\to0$ to a
limit, which will be called the "natural" length of the material curve in
question. The natural measures of material surfaces, volumes, and angles
can be defined in a similar way. It turns out that the natural measure
defined thus constructed for material objects coincides with their measure
in the Riemannian space generated on $V_t$ by the frame field $\check a(x,t)$.

Thus in both cases at any time $t$ at each point of the medium there exists
a frame determining how the measure of 1--3-dimensional material objects at
time $t$ has changed compared with their natural measure. It is this
property of the frame that is important in what follows.

Therefore, let us introduce a notion of elastic strain frame for
elastoplastic media generalizing the corresponding notion for elastic media
and based neither on any concept of actual motion of the medium nor on any
assumptions related to full or partial unloading of parts of the medium.

Suppose that at each time the medium has some "elastic structure", which
permits one to judge about the natural measure near each point of the
medium. More precisely, we shall adopt the following assumption.

{\it Assumption.} For each given time $t$, to each point $x$ of the medium
one can assign a nondegenerate frame $\check a(x,t)$ (which will be called the
elastic strain frame of the medium at the point $(x,t)$) such that the
Riemannian metric generated in $V_t$ by the field $\check a(x,t)$ (the metric of the
space $V_{t,\check a}$) is "natural."

{\it Remark.}

1. The "naturality" of a measure, which is clear if full unloading is
possible, is defined as certain relationships, postulated below, with the
stress and the plastic strain.

2. In a polycrystal material, crystal lattices can serve as a prototype of
an "elastic structure."

The elastic strain frame field on $V_t$ is not unique. Indeed, if $\check a(x,t)$ is
an elastic strain frame field, then for a frame field $\check a^\prime(x,t)$ to be an
elastic strain frame field it is necessary and sufficiently that these
fields be equivalent, i.e., that the frames $\check a(x,t)$ and $\check a^\prime(x,t)$
be equivalent for all $x\in V_t$.

Indeed, if the fields $\check a$ and $\check a^\prime$ are equivalent, then these frames
generate equal scalar products at each point. Consequently, the Riemannian
metrics generated by the fields $\check a$ and $\check a^\prime$ coincide; i.e.,
$\check a^\prime$ also generates a natural metric and hence is an elastic strain frame
field. The necessity can also be easily justified.

The following lemma will be often used in the sequel.

{\it Lemma. For nondegenerate frames $\check a$ and $\check a^\prime$ to be equivalent, it
is necessary and sufficiently that there be an orthogonal $3\times 3$
matrix $Q$ such that}
$$
  \check a^\prime=Q\check a.\eqno(6.5)
$$

Indeed, let the frames be equivalent: $\forall\nkg {\alpha}$, $\nkg {\beta}\quad$
$\langle \nkg {\alpha},\nkg {\beta}\rangle_{\check a^\prime}=\langle\nkg {\alpha},\nkg {\beta}\rangle_{\check a}$.

We denote $Q:=(\check a^{\prime\check a})^T$; then $\check a^\prime=Q\check a$.
Since each of the frames is orthogonal with respect to the scalar product generated
by it, it follows that $\delta_{ij}=\langle\nkg a^\prime_i, \nkg a^\prime_j\rangle_{\check a^\prime}=
\langle \nkg a^\prime_i, \nkg a^\prime_j\rangle_{\check a}=\langle q_{ik}\nkg a_k,q_{jl}\nkg a_l\rangle_{\check a}=
q_{ik}q_{jl}\langle\nkg a_k, \nkg a_l\rangle_{\check a}=q_{ik}q_{jl}\delta_{kl}$
for any $i$ and $j$, i.e., $q_{ik}q_{jk}=\delta_{ij}$ and hence $Q$ is orthogonal. The
sufficiency can be proved even easier.

\section {The plastic strain tensor}

To characterize plastic strain, we need additional information on the
change in the measure of geometric objects undergoing deformation.

{\it 7.1. The measure dilatation coefficients for the mapping of $\mathrm V_{\check\alpha}$ into
$V_{\check\beta}$.} Here the situation is the same as in Sections~4.2--4.3; $\mathrm V_{\check\alpha}$ plays
the role of the domain of the initial position of the medium, $V_{\check\beta}$ plays
the role of the domain~$V_t$ of the final position of the medium, $\check\alpha(x)$ is
used instead of $\check\mathrm a(\mathrm x)$, and the $\check\beta$-measure is used instead of the
geometric measure in $V_t$. We introduce the frame
$$
  \check a(x):=\pd{x(\mathrm x)}{\check\alpha(\mathrm x)},
$$
which plays the role of $\check a(x)$ in 4.2--4.3. Arguing as in 4.2--4.3,    we
arrive at the conclusion that the $\check a$ - measure of objects in $V$
coincides with the $\check\alpha$-measure of their preimages in $\mathrm V$. But then the
following assertion holds.

{\it Assertion. The measure dilatation coefficients comparing the
$\check\alpha$-measure of 1--3-dimen-sional objects in $\mathrm V$ with the $\check\beta$-measure of
their images in $V$ are given by formulas (3.7), (3.9), and (3.16) with
$\check\alpha$ replaced by $\check a$.}

For the $\check a$-measure of the initial angles, using the reasoning and
notation of Section 4.2, we have
$$
  \cos(\widehat{\prg m_1,\prg m}_2)_{\mathrm x,\check\alpha}=     
  \frac{\langle \nkg m_1,\nkg m_2\rangle_{\check a(x)}}
  {|\nkg m_1|_{\check a(x)}\cdot|\nkg m_2|_{\check a(x)}},\eqno(7.1)
$$
where $(\widehat{\bf m_1,\bf m}_2)_{\mathrm x,\check\alpha}$ is the angle between $\bf m_1$
and $\bf m_2$ at the point $\mathrm x$ in the
$\check\alpha$-metric.

{\it 7.2.} In what follows, we consider a medium that moves during the time
interval $[t_b,t_e]=:T$ and is elastoplastic (in the sense of Section 6.2) at each
time $t\in T$. All earlier-introduced assumptions and the main notation are
preserved; we only introduce an additional variable, time $(t)$, which
occurs as an argument of the functions specifying the law of motion of the
medium, the elastic strain frame, the velocities $\nkg v(x,t)$ of points of the
medium, etc.

Let $W:=\{(x,t)|\,t\in T,\,x\in V_t\}$. Let $\check a(x,t)$ be an elastic strain frame
field of the medium of the class $C^1(W)$. We take an arbitrary $t_0\in T$ and denote
the law of motion of the medium starting from time $t_0$ by
$$
  x=x_{t_0}(\mathrm x,t).\eqno(7.2)
$$
To each point $\mathrm x$ in space, this law assigns the spatial position $x$
at time $t\in T$ of the material point residing at $\mathrm x$ at time~$t_0$. We shall
assume that $x_{t_0}\in C^2(V_{t_0}\times T)$.

The comparison of natural measures of material objects at time $t$ and
$t_0$ characterizes the plastic strain of the medium on the interval $[t_0,
t]$.

Just as the frame field $\partial x/\partial\check\mathrm a$ on $V_t$,
where $\check\mathrm a$ defines the geometric
measure at the initial time (since $E=E_{\check\mathrm a}$), completely characterizes the
change in the geometric measure of material objects (see Section 4.3), so
the frame field
$$
  \check b(x,t):=\pd{x_{t_0}(\mathrm x,t)}{\check a(\mathrm x,t_0)},\quad
  (x=x_{t_0}(\mathrm x,t)),\eqno(7.3)
$$
where $\check a(x,t_0)$ defines the natural metric on $V_{t_0}$ at time $t_0$, completely
characterizes the change in the natural measure of material objects in time
$[t_0,t]$, i.e., the plastic strain on this time interval (the elastic strain
frame field $\check a(x,t)$ on $V_t$ and $V_{t_0}$ being known).

Indeed, we take $\forall\mathrm x_0\in V_{t_0}$ and $x_0:=x_{t_0}(\mathrm x_0,t)$ and
momentarily set $\check a(\mathrm x_0,t_0)=:\check a_0=:(\nkg a^0_1,\nkg a^0_2,\nkg a^0_3)^T$,
$\check a(x_0,t)=:\check a$, and $\check b(x_0,t)=:\check b$.
According to the Assertion in Section 7.1 (with $\check a$ as $\check\beta$ and
$\check b$ as $\check a$), the natural length dilatation coefficient at a point $(x_0,t)$ 
in an arbitrary direction $\nkg m$ is equal to
$$
  \tilde K_1(x_0,t)=\frac{|\nkg m|_{\check a}}{|\nkg m|_{\check b}}.\eqno(7.4)
$$

Moreover, for two arbitrary nonzero vectors $\nkg m_1$ and $\nkg m_2$ at $x_0$, one
can take oriented material curves for which these vectors are the positive
tangent vectors at the point $x_0$ at time $t$ and introduce the positive
tangent vectors ${\bf m}_i$ to these curves at time $t_0$ at the point $\mathrm x_0$.

Now, using the properties of the elastic strain frame $\check a$ (see the
assumption in Section~ 6, (4.6), and (7.1) with~$\check\alpha=\check a_0$), for the material
angle $(\widehat{\nkg m_1,\nkg m}_2)_{x_0,t}$ and the corresponding
$(\widehat{{\bf m}_1,{\bf m}}_2)_{\mathrm x_0,t_0}$ we have
$$
\begin{array}{lll}
\cos\left(\nkg m_1\hat ,\nkg m_2\right)_{x_0,t}=
\dfrac{\langle \nkg m_1, \nkg m_2\rangle_{\check a}}
     {|\nkg m_1|_{\check a}|\nkg m_2|_{\check a}}\vspace{.1in}\\
\cos\left({\bf m}_1\hat ,{\bf m}_2\right)_{\mathrm x_0,t_0}=
\dfrac{\langle {\bf m}_1,{\bf m}_2\rangle_{{\it\check a_0}}}
     {|{{\bf m}_1}|_{\check a_0}
      |{\bf m}_2|_{{\it\check a_0}}}=
\dfrac{\langle \nkg m_1, \nkg m_2\rangle_{\check b}}
     {|\nkg m_1|_{\check b}|\nkg m_2|_{\check b}}
\end{array}
\eqno(7.5)
$$

Thus the specification of $\check b$ for known $\check a$ permits one to compute the
natural length dilatation coefficient at $x_0$ at time $t$ in any direction
$\nkg m$ (see (7.4)) and, for any material curves issuing from $x_0$ at time
$t$, compute the cosines of the natural angles between them at the point
$x_0$ at time $t$ and at the point $\mathrm x_0$ at time $t_0$ (see (7.5)).

In particular, for the vectors $\nkg b_i$ of the frame $\check b$, we have
$b^{\check b}_i=\hat i$ and $|\nkg b_i|_{\check b}=1$. Hence (see (7.4)
and (7.5)) $|\nkg b_i|_{\check a}$ is the material length dilatation
coefficient along $\nkg b_i$ at the point $x_0$ at time $t$; moreover,
$\cos(\widehat{\nkg b_i,\nkg b}_j)_{x_0,t}=
\langle \nkg b_i, \nkg b_j\rangle_{\check a}/(|\nkg b_i|_{\check a}|\nkg b_j|_{\check a})$,
while the cosine of the material angle between the corresponding vectors
$\nkg a^0_i$ and $\nkg a^0_j$ of the frame $\check a_0$ at $\mathrm x_0$ at time $t_0$
(see (7.5)) is zero for $i\neq j$.

We see that the frame $\check b$ plays the same role in determining the plastic
strain of the medium in time $[t_0,t]$ as $\check a$ plays in determining the
elastic strain.

By analogy with the Cauchy--Lagrange elastic strain matrix, we introduce
the plastic strain matrix at a point $x_0$ from time $t_0$:
$$
  \hat\varepsilon^p:=\left(\tfrac12(\langle \nkg b_i, \nkg b_j\rangle_{\check a}-\delta_{ij})\right)
  =:\frac{\hat\gamma-1}2.\eqno(7.6)
$$
By introducing the matrix $B:=\check b^{\check a}$ and by taking into account the definition
of the scalar product $\langle\cdot,\cdot\rangle_{\check a}$, we obtain
$$
  \hat\gamma=B^TB.
$$

The physical meaning of $\hat\varepsilon^p$ is the same as that of $\hat\varepsilon^e$
with the only difference that geometric lengths and angles are replaced by natural ones.
Let us study how $\hat\varepsilon^p$ changes if we replace the elastic strain frame
$\check a_0$ at the point $(\mathrm x_0,t_0)$ by the equivalent frame $\check a^\prime _0$.
It suffices keep track of the change in $\hat\gamma$. We denote
$Q_0:=((\check a^\prime _0)^{\check a_0})^T$; then
$$
  \check a^\prime_0=Q_0\check a_0.\eqno(7.7)
$$

For the objects $\check b^\prime$, $B^\prime$, $(\hat\varepsilon^p)^\prime$, and
$\hat\gamma^\prime$ corresponding to $\check a^\prime_0$ to
belong to the class $C^1$, we replace the field $\check a(x,t)$ by an equivalent
field $\check a^\prime(x,t)$ such that
$$
  \check a^\prime(x,t)=Q(x,t)\check a(x,t),\eqno(7.8)
$$
where $Q(x,t)$ is a $C^1$ field of orthogonal matrixes (see the lemma in
Section 6) and $Q(\mathrm x_0,t_0)=Q_0$. Now, in view of (7.3), (7.7) and (7.8), we obtain
$$
  \check b^\prime=\pd{x_{t_0}(\mathrm x_0,t)}{\check a^{\prime}(\mathrm x_0,t_0)}=     
  \pd{x_{t_0}(\mathrm x_0,t)}{\mathrm x}\check a^\prime_0
  =Q_0\check b=Q_0B^T\check a=Q_0B^TQ^T\check a^\prime.
$$
It follows that
$$
  B^\prime=QBQ^T_0.\eqno(7.9)
$$

Therefore, $\hat\gamma^\prime=B^{\prime T}B^\prime=Q_0B^TBQ^T_0=Q_0\hat\gamma Q_0^T$;
i.e., $\hat\gamma$ is the matrix of some covariant rank two
tensor $\nkg {\gamma}$ in the frame $\check a_0$. This tensor is defined at
$(\mathrm x_0,t_0)$ for elastic strain frames and depends on time.

Hence the same it true for $\hat\varepsilon^p$ and for the tensor $\mbox{\boldmath$\varepsilon$}^p$ of plastic strain
at the point $\mathrm x_0$ from time $t_0$.

In general, the present paper deals only with rank two tensors, and since
all of them are associated only with elastic strain frames, it follows that
the matrices of transformation from one frame to another are orthogonal, so
that the tensors are covariant.

{\it Remark.} The tensor $\mbox{\boldmath$\varepsilon$}^p$ is different from the tensor, traditionally
denoted by the same letter, defined as the difference between the full and
elastic strain tensors or reckoned from the initial time.

Let us introduce the plastic strain rate tensor $\nkg {\zeta}(\mathrm x_0,t_0)$ at the
point $(\mathrm x_0,t_0)$. By definition, it is equal to the total $t$-derivative at time
$t_0$ of the plastic strain tensor at the point $\mathrm x_0$ reckoned from time
$t_0$,
$$
  \mbox{\boldmath$\zeta$}(\mathrm x_0,t_0)=
\left({\mathrm d\mbox{\boldmath$\varepsilon$}^p}/{\mathrm dt}\right)_
{|_{t_0}}.\eqno(7.10)
$$

\section { The strain consistency equation}

On the mass point trajectory passing through $\mathrm x_0$ at time $t_0$, we have
$$
  \frac{\partial x_{t_0}(\mathrm x_0,t)}{\partial\mathrm x}
\check a(\mathrm x_0,t_0)=(B(x,t))^T\check a(x,t),\; x=x_{t_0}(\mathrm x_0,t)
\nonumber
$$
according to (7.3). By taking the total  $t$-derivative of this identity at
$t=t_0$ and by changing the order of differentiation on the left-hand
side, we obtain
$$
  \frac{\partial \nkg v(\mathrm x_0,t_0)}{\partial\check a(\mathrm x_0,t_0)}=
\left[\frac{\mathrm d}{\mathrm dt}B^T(x,t)\right]_{|_{t_0}}
\check a(\mathrm x_0,t_0)+B^T(\mathrm x_0,t_0)
\left[\frac{\mathrm d}{\mathrm dt}\check a(x,t)\right]_{|_{t_0}}.
$$
Taking into account the fact that $B^T(\mathrm x_0,t_0)=I$ and passing to contracted notation,
we obtain the equation
$$
  \frac{\partial \nkg v}{\partial \check a}=
\left(\frac{\mathrm d B^T}{\mathrm dt}\right)_{|_{t_0}}\check a+
\left(\frac{\mathrm d \check a}{\mathrm dt}\right)_{|_{t_0}},\eqno(8.1)
$$
satisfied at the point $(\mathrm x_0,t_0)$ by any elastic strain frame field $\check a$ and by
the corresponding $B$ (since we deal with a specific motion of the
medium, the field $\nkg v$ is fixed).

By taking the total $t$-derivative of (7.9), for $t=t_0$ we obtain the
law describing how the matrix $\mathrm dB/d\mathrm t_{|_{t_0}}$ changes under the passage to an
arbitrary elastic strain frame field $\check a^\prime(\mathrm x,t)$ (see (7.8)):
$$
 \frac{\mathrm dB^\prime}{\mathrm dt}_{|_{t_0}}=
\frac{\mathrm dQ}{\mathrm dt}_{|_{t_0}}Q^T_0+
Q_0\frac{\mathrm dB}{\mathrm dt}_{|_{t_0}}Q^T_0.\eqno(8.2)
$$
For this law to be a tensor law on $W$, it is necessary and sufficient that
$$ 
  \frac{\mathrm dQ(\mathrm x_0,t_0)}{\mathrm dt}=0\;\;
\forall (\mathrm x_0,t_0)\in W,\eqno(8.3)    
$$
i.e. that $Q(x,t)$ be constant along the space-time trajectories of the
points of the medium.

We say that elastic strain frame fields $\check a$ and $\check a^\prime$ related by formula
(7.8) with $Q$ satisfying (8.3) are kinematically equivalent.

Thus the set of elastic strain frame fields splits into classes of
kinematically equivalent fields, and the matrix $dB/dt_{|_{t_0}}$
varies according to the tensor law on each of the classes.

These classes are sufficiently ample; namely, the following assertion is
valid: if $\check a(x,t)$ is an elastic strain frame field of the medium in
$W$, then for each $t_0$ and each elastic strain frame field $\check
a_0(x)$ given at time $t_0$ on $V_{t_0}$ there exists a field $\check
a^\ast(x,t)$ kinematically equivalent to $\check a(x,t)$ and coinciding with
$\check a_0(x)$ at time $t_0$.

Indeed, let $Q_0(x)$ be the field of orthogonal matrices such that
$\check a_0(x)=Q_0(x)\check a(x,t_0)$.
We introduce a field of orthogonal matrices on $W$ by setting $Q(x,t_0)=Q_0(x)$
and by requiring that $Q(x,t)$ does not vary along the trajectories of the
points of the medium. Let $\check a^\ast(x,t):=Q(x,t)\check a(x,t)$;
then first, $\check a^\ast(x,t_0):=Q(x,t_0)\check a(x,t_0)=\check a_0(x)$
and second, the field $\check a^\ast$ is kinematically equivalent to $\check a(x,t)$,
since $dQ(x,t)/dt=0$.

{\it Assumption}. For the medium, there exists an elastic strain frame
field of the class $C^1(W)$ for which the matrix $dB/dt_{|_{t_0}}$
is symmetric on $W$. We say that this field is kinematically constrained.

{\it Remark 8.1.}  It is shown in the paper mentioned second in the
footnote in Section~1 that if the law of motion of the medium is of the
class $C^3$, then the existence of a kinematically constrained field
automatically follows from the existence of at least one elastic strain
frame field of the class $C^2$.

It turns out that the set of kinematically constrained elastic strain frame
fields is a class of kinematically equivalent elastic strain frame fields.

Indeed, if $\check a$ is a kinematically constrained elastic strain frame field
and $\check a^\prime$ is a kinematically equivalent field, then
$(dB^\prime/dt)_{|_{t_0}}=Q_0(dB/dt)_{|_{t_0}}Q_0^T$ by (8.2), and
hence this matrix is symmetric (since so is $dB/dt_{|_{t_0}}$);
thus $\check a^\prime$ is also
kinematically constrained. On the other hand, if elastic strain frame
fields $\check a$ and $\check a^\prime$ are kinematically constrained, then the matrices
$dB/dt_{|_{t_0}}$ and $dB^\prime/dt_{|_{t_0}}$ are symmetric, and hence so is
$\left(dQ/dt_{|_{t_0}}\right)Q^T_0$ (see (8.2)).
Consequently, $(dQ/dt)_{|_{t_0}}Q^T_0=
Q_0(dQ^T/dt)_{|_{t_0}}$, and hence
$$
  \frac{\mathrm dQ}{\mathrm dt}_{|_{t_0}}Q^T_0=\frac{1}{2}
\left(\frac{\mathrm dQ}{\mathrm dt}_{|_{t_0}}Q^T_0+
      Q_0\frac{\mathrm dQ^T}{\mathrm dt}_{|_{t_0}}\right)=
\frac{1}{2}\frac{\mathrm d(QQ^T)}{\mathrm dt}_{|_{t_0}}=0,
$$
since $QQ^T=I$. Thus $dQ/dt_{|_{t_0}}=0$; i.e., $\check a$ and $\check a^\prime$
are kinematically equivalent.

The class of kinematically constrained elastic strain frame fields plays a
special role. Let $\check a$ be a kinematically constrained elastic strain frame
field, and let $(\mathrm x_0,t_0)$ be an arbitrary point in $W$. Taking into account the
symmetry of $dB/dt_{|_{t_0}}$, the relation $B_{|_{t_0}}=I$, and formulas (7.10) and (7.6), we
obtain the following expression for the matrix $ \nkg {\zeta}(\mathrm x_0,t_0)$ in the frame $\check a_0$:
\begin{eqnarray*}
\zeta_{\check a_0}(\mathrm x_0,t_0)=
\left[\left(\frac{\mathrm d}{\mathrm dt}
                  \mbox{\boldmath$\varepsilon$}^p\right)_{|_{t_0}}\right]
_{\check a_0}=\left(\frac{\mathrm d}{\mathrm dt}\hat \varepsilon^p\right)_{|_{t_0}}
=\frac{1}{2}
\left[\frac{\mathrm d}{\mathrm dt}(B^TB-I)\right]_{|_{t_0}}=\\=\frac{1}{2}
\left[\left(\frac{\mathrm dB^T}{\mathrm dt}B\right)_{|_{t_0}}+
\left(B^T\frac{\mathrm dB}{\mathrm dt}\right)_{|_{t_0}}\right]=
\frac{\mathrm dB^T}{\mathrm dt}_{|_{t_0}}
\end{eqnarray*}
Substituting this into (8.1), we obtain the strain rate consistency
equation
$$
  \frac{\partial \nkg v}{\partial \check a}=\zeta_{\check a}\check a+
\frac{\mathrm d\check a}{\mathrm dt},\eqno{(8.4)}
$$
which is satisfied in $W$ by each kinematically constrained elastic strain
frame field.

The physical meaning of Eq.~(8.4) becomes clear if we remember how (8.1)
was obtained: the material rate $\partial\nkg v(x,t)/\partial\check a(x,t)$
due to the motion of the medium of the frame coinciding with $\check a(x,t)$
at the point $(x,t)$ at time $t$ is equal
to the material rate $d\check a(x,t)/dt$ of the elastic strain frame plus a correction
term depending on the matrix of the plastic strain rate tensor in the frame
$\check a$.

{\it Remark 8.2.} For "small" elastic strains, more precisely, under the
assumption that there exists an orthonormal frame $\check e$ and a number
$\delta<<1$ such that $|\nkg a_i-\nkg e_i|$ and $|\partial \nkg a_i/\partial x_j|$,
$|\partial \nkg a_i/\partial t|$, $|\partial\nkg v/\partial x_i|\leq\delta$,
it follows from (8.1), neglecting quantities of the order
of $\delta^2$, that
$$
  \frac{\mathrm d}{\mathrm dt}\mbox{\boldmath$\varepsilon$}_{t_0}(x_0,t_0)=
\frac{\mathrm d}{\mathrm dt}\mbox{\boldmath$\varepsilon$}_{t_0}^p(x_0,t_0)+
\frac{\mathrm d}{\mathrm dt}\mbox{\boldmath$\varepsilon$}^e(x_0,t_0),\;\;
\forall(x_0,t_0)\in W,
$$
where the full and plastic strains are reckoned from time $t_0$. It is this
equation that essentially substitutes the traditional equation
$\quad \mathrm d\mbox{\boldmath$\varepsilon$}=
   \mathrm d\mbox{\boldmath$\varepsilon$}^p+
   \mathrm d\mbox{\boldmath$\varepsilon$}^e$.
(Although this substitution is not justified, it leads to correct results
in our situation.)

\section {The elastic and plastic constitutive equations for \\
elastoplastic media "homogeneous and isotropic\\ in the natural state"}

We understand the media specified in the title as media possessing the
following properties.

{\it Assumption 9.1}. Under elastoplastic deformation, the elastic strain
frames and the natural stresses are related by the same elastic
constitutive equation (6.3)$\equiv$(6.4) as in the case of elastic media
homogeneous and isotropic in the natural state.

Let $\check a^\ast$ be the principal elastic strain frame at a point $(x,t)\in W$;
then, by Assumption 9.1, the principal frame
$\check t(\check a^\ast)=(\nkg t_1(\check a^\ast),\nkg t_2(\check a^\ast),\nkg t_3(\check a^\ast))^T$
of natural stresses on the faces
of $\check a^\ast$ and the frame $\check a^\ast$ are coaxial; here
$\nkg t_i(\check a^\ast)$ are the principal
natural stresses. The only possible nonzero coordinate $\nkg t_i(\check a^\ast)$
in the frame $\check a^\ast $ is $t_{\underline{ii}}(\check a^\ast)=:t_i(\check a^\ast)$
will also be referred to as the principal stress, which, in view of our
notation, will not lead to a misunderstanding;
$(t_1(\check a^\ast),t_2(\check a^\ast),t_3(\check a^\ast))=:\mathrm t(\check a^\ast)$.

{\it Assumption 9.2}. If $\check a^\ast$ is the principal elastic strain frame of the
medium at a point $(x,t)$, then the tensor $\nkg {\zeta}(x,t)$ is coaxial to $\check a^\ast$
for all $(x,t)\in W$; moreover, there exist known functions $p_i:\,{\bf R}^3\to\bf R$ ($i=1,2,3$),
describing the plastic properties of the medium such that
$$
 \zeta_{\check a^{\ast}}(x,t)=
\left|\left|\begin{array}{ccc}
p_1(\mathrm t(\check a^{\ast}))&0&0\\
0&p_2(\mathrm t(\check a^{\ast}))&0\\
0&0&p_3(\mathrm t(\check a^{\ast}))\\
\end{array}\right|\right|=:P(\mathrm t(\check a^{\ast})) .\eqno(9.1)
$$

{\it Remark.} 1. Thus the medium in question is so far "perfectly plastic"
in the sense that the plastic strain rates are independent of the strain
history and depend only on the principal stresses.

2. The assumption that $\zeta_{\check a^\ast}$ depends only on the principal stresses and is
independent of the stress increments seems to us to be quite natural for
"slow" deformations.

3. Using the technique applied in the study of the yield locus of an
isotropic material (e.g., see [14]), one can readily show that
$$
  p_3(t_2,t_3,t_1)=p_2(t_3,t_1,t_2)=p_1(t_1,t_2,t_3)=:p(\mathrm t)\quad
  \forall\mathrm t.
$$

For an arbitrary elastic strain frame $\check a$ at a point $(x,t)$, by $R$ we
denote the orthogonal matrix such that $\check a=R\check a^\ast$.
Then $\zeta_{\check a}=R\zeta_{\check a^\ast}R^T$, and the plastic
constitutive equation of the medium (9.1) can be represented in the form
$$
  \zeta_{\check a}(x,t)=RP(\mathrm t(\check a^\ast))R^T\quad
  \forall(x,t)\in W.\eqno(9.2)
$$

\section {The system of elastoplastic strain equations in the Eulerian
variables}

Let us summarize our study. We consider an elastoplastic medium satisfying
the assumptions in Sections 6 and 8 and moving in a space-time domain $W$.
The natural density $\nkg {\Phi}(x,t)$ of applied mass forces of the class $C(W)$ and the
density $\rho_0$ of the undeformed medium are known. The unknowns are the
velocity field $\nkg v(x,t)$ of the points of the medium and any of the kinematically
equivalent kinematically constrained elastic strain frame fields $\check a(x,t)$ of
the medium, where~$\check a$ and $\nkg v$ are related by two equations, namely, the
strain consistency equation (8.4) (or (8.1) is $\check a$ not kinematically
constrained) and the equation of motion obtained from the force equilibrium
equation (5.6) by the substitution of $\check a(x,t)$ for $\check b(x)$. Here the relative
stresses $\nkg u_i(x)$ become the natural stresses $\nkg t_i(x,t)$, and the mass forces
include also inertial forces; i.e., $\nkg {\Psi(x)}$ is replaced by
$\nkg {\Phi}(x,t)-\rho_0(d\nkg v/dt)$.  

These equations are valid for any media with an "elastic structure." Along
with the unknowns $\check a$ and $\nkg v$, they also contain the natural stress
frame $\check t$ and a matrix describing the plastic strain; therefore, to
obtain a closed system, one should supplement these equations with
equations describing the elastic and plastic properties of the material.

For media homogeneous and isotropic in the undeformed state, i.e.,
satisfying Assumptions 9.1 and 9.2, the supplementing equations are the
elastic and plastic constitutive equations (6.4) and (9.2).

Thus the system of dynamic equations of elastoplastic strain in this case
has the form
$$
\begin{array}{ll}     
\rho_0\dfrac{\mathrm d \nkg v}{\mathrm dt}=
\dfrac{\partial \nkg t_i}{\partial \nkg a_i}+
\left(\mathrm{div}\nkg a_i-
\dfrac{\partial\ln|\check a|}{\partial \nkg a_i}\right) \nkg t_i+\nkg {\Phi}&\,\\
\dfrac{\mathrm d\check a}{\mathrm dt}=
\dfrac{\partial \nkg v}{\partial \check a}-RPR^T\check a,\;\;\forall(x,t)\in W
&\;
\end{array}
\eqno(10.1)
$$ 
 
$$
\nkg t_i=f_{ji}(\hat\varepsilon^e(\check a))\nkg a_j .\eqno(10.2)
$$

The total derivatives in (10.1) are expressed via $\nkg v$ and $\check a$ as
follows:
$$
\frac{\mathrm d \nkg v}{\mathrm dt}=\frac{\partial \nkg v}{\partial \nkg v}+
                                \frac{\partial \nkg v}{\partial t},\;
\frac{\mathrm d\check a}{\mathrm dt}=\frac{\partial\check a}{\partial \nkg v}+
                                  \frac{\partial\check a}{\partial t}.\eqno(10.3)
$$

System (10.1) (in view of (10.2)) is a system of a vector equation and a
frame equation for the vector function $\nkg v$ and the frame function $\check a$.
In vector form, it consists of four equations for $\nkg v$ and $\nkg a_i$ ($i=1,2,3$).

The equations are represented in coordinate-free form. After the
introduction of some coordinate system and the replacement of derivatives
along vectors by their expressions via partial derivative with respect to
coordinates, one obtains a system of quasilinear first-order partial
differential equations.

Equations (10.1) are solved for the time derivatives of the desired
functions, which makes the system suitable for numerical time-marching
solution.

If the functions $p_i$ describing the plastic properties of the medium
depend not only on the principal stresses but also on parameters determined
by the strain history, then one should supplement the system with equations
relating these parameters to the desired functions.

In the absence of plastic strains ($p_i=0$), system (10.1) passes into system
(41) of elasticity equations.\footnote {See Solomeshch I.A. and
Solomeshch M.A., Elasticity Equations for the Strain Frame and the
Velocities of Points at the an Arbitrary Initial State of the Medium [in Russian],
No.~1941-B93, VINITI, Moscow, 1993.}


\begin{thebibliography}{14}

\bibitem {[1]}  I.~A.~Solomeshch and M.~A.~Solomeshch, ``The elastoplasticity
equations for the elastic strain frame and the velocities of points,''
Dokl. RAN, Vol.~354, No.~6, pp.~759--761, 1997.


\bibitem {[2]} P.~M.~Naghdi, ``A Critical review of the state of finite
plasticity,'' ZAMP, Vol.~41, No.~3, pp.~315--394, 1990.

\bibitem {[3]} A.~S.~Khan and S.~Huang, Continuum Theory of Plastisity, Wiley, New
York, 1995.

\bibitem {[4]} A.~Reuss, ``Ber\"ucksichtigung der elastischen Form\"anderung in
der Plastizit\"atstheorie,'' ZAMM, Vol.~10, No.~3, pp.~266--274, 1930.

\bibitem {[5]} A.~E.~Green and P.~M.~Naghdi, ``A general theory of an
elastic--plastic continuum,'' Arch. Ration. Mech. Anal., Vol.~18, No.~4,
pp.~251--281, 1965.

\bibitem {[6]} M.~B.~Rubin, ``Plasticity theory formulated in terms of physically
based microstructural variables,'' Intern. J. Solids Structures. 1994. V.
31.19. P. 2615-2634.

\bibitem {[7]} J.~F.~Besseling and E. van der Giessen, Mathematical Modelling of
Inelastic Deformation, Chapman and Hall, London, 1994.

\bibitem {[8]} C.~Eckart, ``The thermodynamics of irreversible processes. IV. The
theory of elasticity and plasticity,'' Phys. Rev., Vol.~73, No.~4,
pp.~373--382, 1948.

\bibitem {[9]} P.~M.~Naghdi and A.~R.~Srinivasa, ``A dynamical theory of structured
solids. I. Basic developments,'' Phil. Trans. Roy. Soc. London. Ser. A,
Vol.~345, No.~1677, pp.~425--458, 1993.

\bibitem {[10]} E.~H.~Lee and D.~T.~Liu, ``Finite-strain elastic-plastic theory
with application to plane-wave analysis,'' J. Appl. Phys., Vol.~38, No.~1,
pp.~19--27, 1967.

\bibitem {[11]} A.~V.~Shitikov, ``On a variational principle for the construction
of the elastoplasticity equations at finite strains,'' PMM [Applied
Mathematics and Mechanics], Vol.~59, No.~1, pp.~158--161, 1995.

\bibitem {[12]} P.~Ciarlet, Mathematical Elasticity [Russian translation], Mir,
Moscow, 1992.

\bibitem {[13]} A.~A.~Il'yushin, Continuum Mechanics, Izd-vo MGU, Moscow, 1990.

\bibitem {[14]} R.~Hill, The Mathematical Theory of Plasticity [Russian
translation], Gostekhizdat, Moscow, 1956.
\end{thebibliography}
\end{document}